\documentclass[twocolumn,groupedaddress,amssymb,epsfig,showpacs,10pt]{revtex4}
\usepackage{float}
\usepackage{graphics,epsfig}
\usepackage{graphicx}
\usepackage{dcolumn}
\usepackage{bm}
\usepackage{CJK}
\newcommand{\newc}{\newcommand}

\newc{\be}{\begin{equation}}
\newc{\ee}{\end{equation}}
\newc{\ba}{\begin{eqnarray}}
\newc{\ea}{\end{eqnarray}}
\newc{\bea}{\begin{eqnarray*}}
\newc{\eea}{\end{eqnarray*}}
\newc{\D}{\partial}
\newc{\ie}{{\it i.e.} }
\newc{\eg}{{\it e.g.} }
\newc{\etc}{{\it etc.} }
\newc{\etal}{{\it et al.}}

\newc{\ra}{\rightarrow}
\newc{\lra}{\leftrightarrow}
\newc{\lsim}{\buildrel{<}\over{\sim}}
\newc{\gsim}{\buildrel{>}\over{\sim}}

\begin{document}
\baselineskip=0.8 cm
\title{{\bf Approximate $w_\phi\sim\Omega_\phi$ Relations in Quintessence Models}}

\author{Mingxing Luo$^{1}$}
\email{luo@zimp.zju.edu.cn}

\author{Qi-Ping Su$^{1,2}$}
\email{sqp@itp.ac.cn}

\affiliation{
    $^{1}$ Zhejiang Institute of
Modern Physics, Department of Physics, Zhejiang University,
Hangzhou, Zhejiang 310027, P R China
\\ $^{2}$ Key Laboratory of Frontiers in Theoretical Physics, Institute of
Theoretical Physics, Chinese Academy of Sciences, P.O. Box 2735,
Beijing 100190, China}

\vspace*{0.2cm}
\begin{abstract}
\baselineskip=0.6 cm

Quintessence field is a widely-studied candidate of dark energy.
There is ``tracker solution'' in quintessence models,
in which evolution of the field $\phi$ at present times is not sensitive to its initial conditions.
When the energy density of dark energy is neglectable~($\Omega_\phi\ll1$),
evolution of the tracker solution can be well analysed from ``tracker equation''.
In this paper, we try to study evolution of the quintessence field from ``full tracker equation'',
which is valid for all spans of $\Omega_\phi$.
We get stable fixed points
of $w_\phi$ and $\Omega_\phi$ (noted as $\widehat w_\phi$ and $\widehat\Omega_\phi$) from the ``full tracker equation'', i.e., $w_\phi$ and $\Omega_\phi$ will always approach $\widehat w_\phi$ and $\widehat\Omega_\phi$ respectively.
Since $\widehat w_\phi$ and $\widehat\Omega_\phi$ are analytic functions of $\phi$,
analytic relation of $\widehat w_\phi\sim\widehat\Omega_\phi$ can be obtained, which is
a good approximation for the $w_\phi\sim\Omega_\phi$ relation and can be obtained for the most type of quintessence potentials.
By using this approximation, we find that inequalities $\widehat w_\phi<w_\phi$ and $\widehat\Omega_\phi<\Omega_\phi$ are statisfied if the $w_\phi$ (or $\widehat w_\phi$) is decreasing with time.
In this way, the potential $U(\phi)$ can be constrained directly from observations,
by no need of solving the equations of motion numerically.
\end{abstract}
\pacs{95.36.+x, 98.80.-k, 98.80.Es\\
Key words: quintessence, tracker equation, approximation}
\maketitle

\section{Introduction}

Present astronomical observations require the existence of dark energy,
a significant component of the universe with a negative pressure
\cite{Riess:1998cb,Riess:2004nr,Riess:2006fw,Spergel:2003cb,Komatsu:2008hk}.
Though it has been more than ten years since its discovery,
one is yet to tell what the dark energy is.
We are still analyzing properties of dark energy from observational data and seeking suitable candidates.
Most properties of dark energy depend on two parameters:
the equation of state $w_{de}$ and the fractional energy density $\Omega_{de}$.
Once the $w_{de}\sim\Omega_{de}$ relation is obtained,
we know almost all we need.
At present, it is still not possible to constrain the evolution of dark energy from observations
\cite{Daly:2004gf,Mignone:2007tj,Sullivan:2007pd}.
There are only definite constraints of present values of $w_{de}$ and $\Omega_{de}$ from observations:
$w_{de}^{(0)}$ is rather close to $-1$ and $\Omega_{de}^{(0)}$ is dominating (about 70\%)
\cite{Lazkoz:2007zk,Copeland:2006wr,Mantz:2007qh}.
More constraints on $w_{de}$ and $\Omega_{de}$ will be forthcoming from future observations,
to get the evolution of the $w_{de}\sim\Omega_{de}$ relation from the observations,
more theoretical efforts should be made.

At present, the most economical candidate of dark energy is still the cosmological constant $\Lambda$,
whose equation of state $w_\Lambda=-1$.
There is only a free parameter $\Omega_\Lambda$ in the flat $\Lambda$CDM model.
But it suffers from several problems,
such as the coincidence problem and the fine tuning problem.
Another well studied candidate is the quintessence $\phi$,
a slowly rolling scalar field, analogous to the inflaton.
Its equation of state is $w_\phi=(\dot{\phi}^2/2-U)/(\dot{\phi}^2/2+U)$
so one has $-1\leq w_\phi\leq1$.
In quintessence models, the coincidence problem and the fine tuning problem can be alleviated \cite{Copeland:2006wr}.
For example,
there are tracker solutions for certain type of quintessence models,
in which the evolution of $\phi$ today is not sensitive to its initial conditions at early times \cite{Steinhardt:1999nw}.
The coincidence problem thus becomes less severe.
But it is difficult to find quintessence models with analytic solutions of equation-of-motion,
due to the existence of background matters (dark matter, baryon and radiations).
To study evolutions of quintessence models and to be compared with observations,
one usually has to solve the equations numerically.
There are efforts to find analytic approximations for solutions of equations of motions,
such as \cite{Watson:2003kk} which gives a first order approximation solution for inverse power law potentials.

In this paper, we will try to approximate the $w_\phi\sim\Omega_\phi$
relation at the recent $\Omega_\phi$ dominating period in a semi-analytic way.
To make sure
that the evolution of $\phi$ at present only depends on $U(\phi)$, we
assume there was tracking solution at early times. In
\cite{Steinhardt:1999nw}, conditions for the existence of tracker
solution was given by the ``tracker equation'', which is a
differential equation for $w_\phi$. But this ``tracker equation'' are
only valid as $\Omega_\phi\ll1$. For our purpose, we need a full
tracker equation that is valid for all $\Omega_\phi$ without
conditions attached. Such an equation has been obtained
\cite{Scherrer:2005je,Chiba:2005tj,Lee:2006gx} and will be used here
to study evolutions of quintessence models.

The paper is organized as follows.
In section II, we introduce two new functions $\widehat w_\phi$ and $\widehat \Omega_\phi$
which are fixed points of the full tracker equation.
Assuming that $\Gamma\equiv U''U/U'^2$ and $\epsilon\equiv (U'/U)^2/2$ are nearly constant,
we find that the fixed points are stable for $w_\phi$ and $\Omega_\phi$ if $\Gamma\geq1$.
If $\Gamma$ and $\epsilon$ do not evolve extremely fast,
the relation of $w_\phi\sim\Omega_\phi$ will always approach to that of $\widehat w_\phi\sim\widehat \Omega_\phi$.
In section III we show comparisons between \{$\widehat w_\phi(\phi)$, $\widehat \Omega_\phi(\phi)$\}
and \{$w_\phi(\phi)$, $\Omega_\phi(\phi)$\} numerically for several typical quintessence models.
The relation of $\widehat w_\phi\sim\widehat \Omega_\phi$ is shown to be a good approximation for the
$w_\phi\sim\Omega_\phi$ relation.
In section IV we show how to constrain $U(\phi)$ directly from observational conditions on $w_\phi$ and $\Omega_\phi$
through $\widehat w_\phi$ and $\widehat \Omega_\phi$.
Observational conditions are converted to simple inequalities for $U(\phi)$.
We conclude in section V with discussions.

\section{Get the approximation of $w_\phi\sim\Omega_\phi$ relations}

The equations of motion for quintessence field are
\ba
&\ddot\phi+3H\dot\phi+U'=0&\nonumber\\
&H^2\equiv(\frac{\dot a}{a})^2=\frac{1}{3}(\rho_m+\rho_r+\frac{1}{2}\dot{\phi}^2+U)&
\ea
from which one gets the equations for $w_\phi$ and $\Omega_\phi$:
\ba
\epsilon&=&\frac{3(1+\omega_\phi)}{2\Omega_\phi}(1+\frac{\dot{x}}{6})^2\label{e2} \\
\Gamma-1&=&\frac{\omega_b-\omega_\phi}{2(1+\omega_\phi)}
-\frac{2}{(1+\omega_\phi)}\frac{\ddot{x}}{(6+\dot{x})^2}
-\frac{1+\omega_b-2\omega_\phi}{2(1+\omega_\phi)}\frac{\dot{x}}{6+\dot{x}}\nonumber\\
&&-\frac{3(\omega_b-\omega_\phi)}{(1+\omega_\phi)(6+\dot{x})}\Omega_\phi
\label{e3} \ea where
$$x\equiv\frac{1+\omega_\phi}{1-\omega_\phi}=\frac{1}{2}{\dot{\phi}^2\over U}, ~~~
\dot{x}\equiv {d\ln{x} \over d\ln{a}}, ~~~ \ddot{x}\equiv
{d^2\ln{x} \over d\ln{a}^2}$$ and $a$ is the expansion factor.
We have assumed a flat universe ($\Omega_{b}+\Omega_{\phi}=1$) and set $M_{pl}\equiv1/\sqrt{8\pi G}=1$.
The subscript $b$ represents the dominating background matter.
As $\Omega_\phi\ll1$ at early times, Eq.(\ref{e3}) reduces to the ``tracker equation'' in \cite{Steinhardt:1999nw}.
At the recent acceleration era,
$\Omega_\phi$ is dominating and can not be neglected.
One must use the full tracker equation Eq.(\ref{e3}).
Note also $w_b=0$ in this case.
In this paper, we assume that there was a long enough tracking period at early times,
so that the evolution of the field at present depends only on $U(\phi)$.

Eliminating $\Omega_\phi$ in Eq.(\ref{e3}) by using Eq.(\ref{e2}), one gets:
\ba
\Gamma-1&=&-\frac{\omega_\phi}{2(1+\omega_\phi)}
-\frac{2}{(1+\omega_\phi)}\frac{\ddot{x}}{(6+\dot{x})^2}\nonumber\\
&&-\frac{1-2\omega_\phi}{2(1+\omega_\phi)}\frac{\dot{x}}{6+\dot{x}}
+\frac{\omega_\phi}{8\epsilon}(6+\dot x)
\label{e4}
\ea
For constant $\epsilon$ and $\Gamma$, the fixed point (also called critical point) of Eq.(\ref{e4})
(obtained by setting $\dot x=0$ and $\ddot x=0$):
\be
\widehat{\omega}_\phi=
\frac{1}{6}\left(-3-2\epsilon+4\epsilon\Gamma-\sqrt{(3-2\epsilon+4\epsilon\Gamma)^2-24\epsilon}\right)
\label{e6}
\ee
is stable only if
\be
\Gamma\geq\frac{5+3\widehat\Omega_\phi}{6+2\widehat\Omega_\phi}
\label{e5}
\ee
where the $\widehat\Omega_\phi$ value of the fix point is obtained from Eq.(\ref{e6}) and (\ref{e2})
(also setting $\dot x=0$):
\be
\widehat{\Omega}_\phi=
\frac{1}{4\epsilon}\left(3-2\epsilon+4\epsilon\Gamma-\sqrt{(3-2\epsilon+4\epsilon\Gamma)^2-24\epsilon}\right)
\label{e7}
\ee
When Eq.(\ref{e5}) is satisfied, $\widehat\Omega_\phi$ is also stable.
In this case, ${\omega}_\phi$ and $\Omega_\phi$ will always approach $\widehat{\omega}_\phi$ and $\widehat\Omega_\phi$ respectively.
In this paper, we will only study the case of $\Gamma\geq1~~(i.e.,~w_\phi\leq w_b)$,
so Eq.(\ref{e5}) is guaranteed for all spans of $\widehat\Omega_\phi$.

$\Gamma$ and $\epsilon$ generally are not constants, as they are functions of $U(\phi)$.
The above results are still valid if the evolution of $\widehat{w}_\phi$ is not extremely fast,
which can be satisfied in the most quintessence models.
In this case, $w_\phi$ and ${\Omega}_\phi$ will keep on chasing the dynamic $\widehat{w}_\phi$ and $\widehat{\Omega}_\phi$.

Giving the form of $U(\phi)$ of a quintessence model,
one gets parametric functions $\widehat{w}_\phi(\phi)$ and $\widehat{\Omega}_\phi(\phi)$
from Eq.(\ref{e6}) and (\ref{e7}),
and thus the analytic relation of $\widehat{w}_\phi\sim\widehat{\Omega}_\phi$.
For certain models, there are simple and explicit relations of $\widehat{w}_\phi\sim\widehat{\Omega}_\phi$.
For example, for power law potentials $U=U_0/\phi^n~~(n>0)$ one has:
\be
\widehat{w}_\phi=-\frac{1}{1+n(1-\widehat{\Omega}_\phi)/2}
\ee
The $\widehat{w}_\phi\sim\widehat{\Omega}_\phi$ relation is a good approximation for that of
$w_\phi\sim{\Omega}_\phi$, as the evolution of $w_\phi\sim{\Omega}_\phi$ will approach that of $\widehat{w}_\phi\sim\widehat{\Omega}_\phi$.
We will show this in the next section.
In this way, evolutions of quintessence models can be studied directly from $U(\phi)$.

\section{Compared with numerical results}

\begin{figure*}[b!]
\includegraphics[width=14cm,height=8cm]{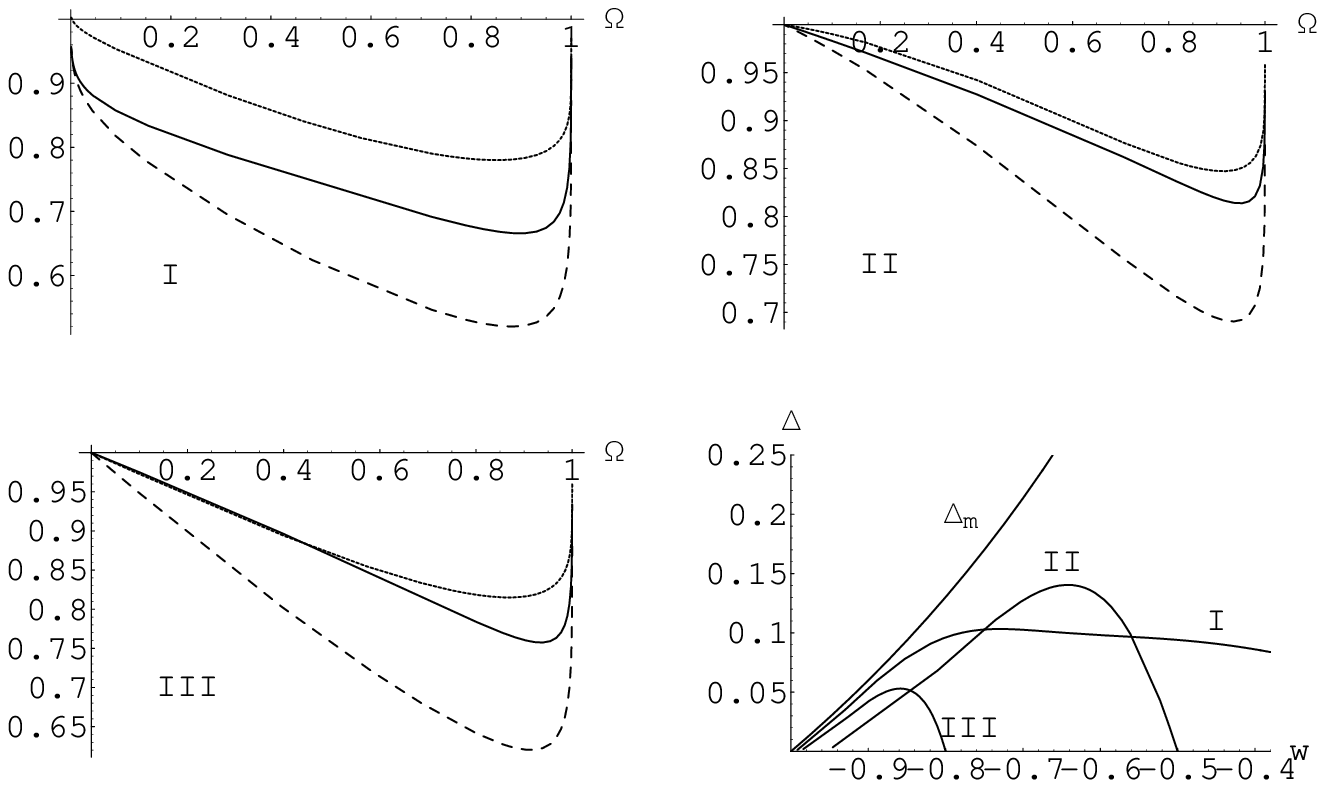}
\caption{Evolution of $(1+\dot{x}/6)^2$ (solid lines),
$(1+\widehat w_\phi)/(1+w_\phi)$ (dashed lines) and
$\widehat\Omega_\phi/\Omega_\phi$ (dotted lines) with respect to $\Omega_\phi$.
The potentials: I. $U=U_0e^{1/\phi}$;
II. $U=U_0/\phi^{2}$; III. $U=U_0/\phi^{0.5}$.
The last figure shows the deviation $\Delta$ of ${\widehat w}_\phi$ from $w_\phi$ for these models.
}
\label{f1}
\end{figure*}

\begin{figure*}[t!]
\includegraphics[width=14cm,height=8cm]{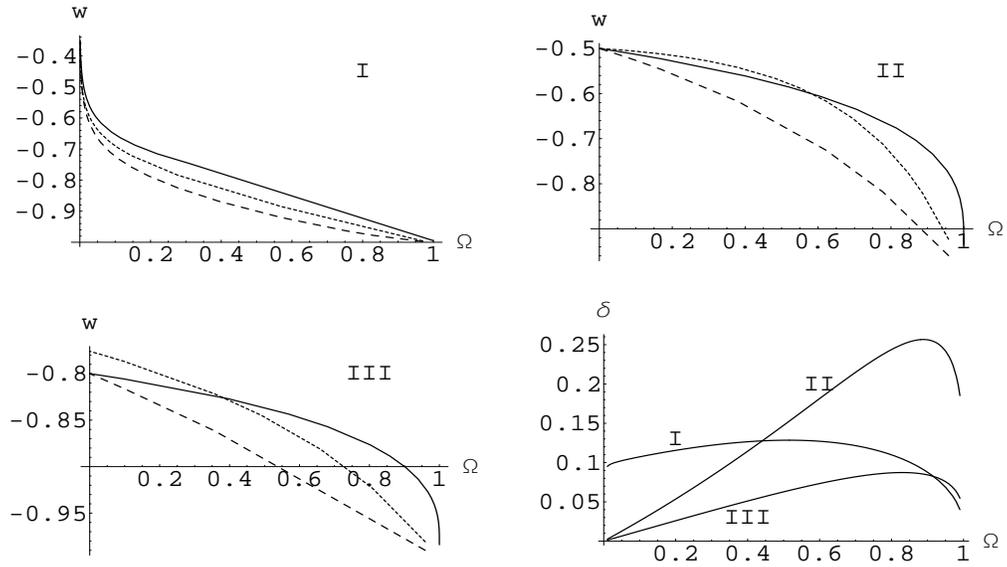}
\caption{The $w_\phi\sim\Omega_\phi$ relation (solid lines),
the $\widehat w_\phi\sim\widehat\Omega_\phi$ relation (dashed lines) and
the $\widetilde w_\phi\sim\widetilde\Omega_\phi$ relation (dotted lines)
in the $w-\Omega$ space.
The potentials: I. $U=U_0e^{1/\phi}$;
II. $U=U_0/\phi^{2}$; III. $U=U_0/\phi^{0.5}$.
The last figure shows the differences between the relation of $w_\phi\sim\Omega_\phi$
and that of $\widehat w_\phi\sim\widehat\Omega_\phi$ for these models
( $\delta=-[ w(\Omega)-\widehat w(\widehat\Omega=\Omega) ]/w(\Omega)$ ).
}
\label{f}
\end{figure*}

In this section we will show that
${\widehat w}_\phi$ and ${\widehat\Omega}_\phi$ are good approximations for $w_\phi$ and $\Omega_\phi$,
and so is the $\widehat{w}_\phi\sim\widehat{\Omega}_\phi$ relation for that of $w_\phi\sim{\Omega}_\phi$.
We have checked it for a variety type of quintessence potentials,
and typical examples are shown in Fig. \ref{f1} and Fig. \ref{f}.
The accuracy of this approximation is precise enough to study the evolution properties of quintessence models,
especially the models that are favored by present observations.
As $w_\phi$ must decrease from its tracking value (close to $w_b$) to present value (close to $-1$),
we will only study models in which $w_\phi$ decreases monotonously ($\dot{x}<0$).

At first we estimate differences between $\widehat{w}_\phi$ and $w_\phi$
and between $\widehat{\Omega}_\phi$ and ${\Omega}_\phi$.
The Eq.(\ref{e2}) can be rewritten as:
\ba
\frac{1+\widehat w_\phi}{\widehat\Omega_\phi}=\frac{1+\omega_\phi}{\Omega_\phi}(1+\frac{\dot{x}}{6})^2&\nonumber\\
\Rightarrow~~~~~(\frac{1+\widehat w_\phi}{1+w_\phi})\cdot(\frac{\widehat\Omega_\phi}{\Omega_\phi})^{-1}=(1+\frac{\dot{x}}{6})^2
\label{t1}
\ea
For a variety of quintessence models, we have seen numerically that $(1+\widehat w_\phi)/(1+w_\phi)$ and
$\widehat\Omega_\phi/\Omega_\phi$ have the similar evolving forms as that of $(1+\dot{x}/6)^2$ and
$1>\widehat\Omega_\phi/\Omega_\phi\gtrsim(1+\dot{x}/6)^2>(1+\widehat w_\phi)/(1+w_\phi)$.
Typical examples are shown in Fig. \ref{f1}.
If the evolution of $\widehat w_\phi$ (and $w_\phi$) is slower, the value of $\dot{x}$ will be closer to $0$,
and the differences between $w_\phi,~\Omega_\phi$ and their fixed points will be smaller.

There is a lower bound $\dot{x}>6w_\phi/(1-2w_\phi)$ given in \cite{Scherrer:2005je,Chiba:2005tj}.
As $\widehat\Omega_\phi/\Omega_\phi$ is much closer to $1$ compared with $(1+\widehat w_\phi)/(1+w_\phi)$,
one gets a upper bound for the deviation $\Delta$ of ${\widehat w}_\phi$ from $w_\phi$
by setting $\widehat\Omega_\phi/\Omega_\phi\simeq1$
in Eq.(\ref{t1}):
\be
\Delta\equiv\frac{\widehat w_\phi-w_\phi}{w_\phi}\lesssim\Delta_m=\frac{(2-3w_\phi)(1+w_\phi)}{(1-2w_\phi)^2}
\ee
which is rather small when $w_\phi$ is close to $-1$, as shown in Fig. \ref{f1}.
Present observations indicate that $w_\phi$ is rather close to $-1$ at low redshift.
For most models $\Delta$ is much smaller than this bound as $w_\phi$ is not so close to $-1$,
as shown in Fig. \ref{f1}.
The deviation of $\widehat\Omega_\phi$ from $\Omega_\phi$ is also small.

The $\widehat{w}_\phi\sim\widehat{\Omega}_\phi$ relation thus is a good approximation
for the $w_\phi\sim{\Omega}_\phi$ relation.
Several examples are shown in Fig. \ref{f}.
At the early tracking era, $\widehat\Omega_\phi<<1$ and
the relation of $w_\phi\sim\Omega_\phi$ is almost the same as that of
$\widehat w_\phi\sim\widehat\Omega_\phi$.
When $\widehat\Omega_\phi$ becomes unnegligible,
the curve of $\widehat w_\phi\sim\widehat\Omega_\phi$ will begin to get away from that of $w_\phi\sim\Omega_\phi$
in the $w-\Omega$ space.
The curve of $w_\phi\sim\Omega_\phi$ will chase after that of $\widehat w_\phi\sim\widehat\Omega_\phi$.
Normally $\widehat w_\phi$ will tend to $-1$ and $\widehat\Omega_\phi$ will tend to $1$ at last,
and the two curves will be close to each other once again.

Empirically, we have also found a better approximation for the relation of $w_\phi\sim\Omega_\phi$
on the basis of $\widehat w_\phi$ and $\widehat\Omega_\phi$:
\be
\widetilde w_\phi=\widehat w_\phi+\frac{(2+\widehat
w_\phi)\widehat\Omega_\phi-2\widehat w_\phi-1}
{5-3\widehat\Omega_\phi}(1+\widehat w_\phi),~~~~\widetilde \Omega_\phi=\widehat\Omega_\phi
\label{e17}
\ee
The curve of $\widetilde{w}_\phi\sim\widetilde \Omega_\phi$ is much closer to
that of $w_\phi\sim\Omega_\phi$, as shown in Fig. \ref{f}.

\section{Constrain quintessence potentials}

In the above, we have obtained approximations $\widehat w_\phi$ and $\widehat\Omega_\phi$ for $w_\phi$ and $\Omega_\phi$
which are analytic functions of $U(\phi)$.
We will show how to constrain $U(\phi)$ directly from observational results on $w_{de}$ and $\Omega_{de}$
through $\widehat w_\phi$ and $\widehat\Omega_\phi$.
Present data seems to indicate that $w_{de}^{(0)}<-0.8$ and $0.7 \lesssim \Omega_{de}^{(0)}<0.8$
\cite{Lazkoz:2007zk,Mantz:2007qh}.
As more conditions on dark energy to be obtained in future observations,
more quintessence models can be checked with directly by using our method.

At the early tracking era $w_\phi$ was close to $w_r=1/3$ \cite{Bludman:2004az,Zlatev:1998tr},
and present $w_{de}^{(0)}$ is very close to $-1$.
Taking this for guidance, here we consider only quintessence models in which $w_\phi~$(and $\widehat w_\phi$)
keeps on decreasing monotonously ($\dot x<0$).
This is guaranteed if $U(\phi)$ satisfies the equation:
\be
\frac{d\ln{(\Gamma-1)}}{d\ln{U}}<\frac{3}{2\epsilon}(1-\frac{1}{2\Gamma-1})
\ee
In this case, one finds the following inequalities
\be
\widehat w_\phi<w_\phi,~~~~\widehat \Omega_\phi<\Omega_\phi
\label{e12}
\ee
if the evolution of $w_\phi$ is not extremely fast.
Intuitively, the curve of $\widehat w_\phi\sim\widehat \Omega_\phi$ is always on the up side of
that of $w_\phi\sim\Omega_\phi$ in the $w-\Omega$ space, as shown in Fig. \ref{f}.

With the help of inequalities (\ref{e12}),
$U(\phi)$ can be constrained directly from conditions on $(w_{de}, \Omega_{de})$.
Take
\be
(w_{de}^{(0)}<-0.8, \Omega_{de}^{(0)}<0.8)
\label{e14}
\ee
for illustration \cite{Caldwell:2005tm}.
Since $\widehat w_\phi$ decreases monotonously as $\widehat\Omega_\phi$ increases,
we thus have $\widehat w_\phi(\widehat\Omega_\phi=0.8)<-0.8$.
This inequality can be converted to:
\be
\Gamma(\epsilon=3/8)>7/5
\label{e15}
\ee
which is a necessary condition for inequalities (\ref{e14}).

\begin{table*}[t!]
\begin{minipage}{0.88\textwidth}
\caption{Constraints of typical potentials of quintessence}
\end{minipage}\\
\begin{tabular}{ccccc}
\hline
\hline
\bf{$U(\phi)~(n>0,\phi>0)$} &~~~  $\epsilon\equiv\frac{1}{2}(\frac{U'}{U})^2$ & $\Gamma\equiv \frac{U''U}{U'^2}$ &
~$\Gamma(\epsilon=\frac{3}{8})>\frac{7}{5}$~
&~$\Gamma(\epsilon=\frac{3}{28})<\frac{77}{20}$~
\\\hline \vspace{-5pt}
$\frac{U_0}{\phi^n}$ & $\frac{n^2}{2\phi^2}$ & $1+\frac{1}{n}$ & $n<\frac{5}{2}$ &
 $n>\frac{20}{57}$
\\
$U_0e^{n/\phi}$&$\frac{n^2}{2\phi^4}$&$1+\frac{2\phi}{n}$
 & $n<\frac{50}{\sqrt{3}}$ & $n>1$
\\
$\frac{U_0}{\phi^n}e^{\phi^2/2}$ & ~~$\frac{(n-\phi^2)^2}{2\phi^2}$~~ &~~ $1+\frac{(n+\phi^2)}{(n-\phi^2)^2}$ ~~& $n>0$
& $\emptyset$
\\
\hline \hline
\end{tabular}
\end{table*}

If $w_\phi$ is too close to $-1$,
it will be difficult to distinguish quintessence models from the cosmological constant \cite{Caldwell:2005tm}.
Take $w_{\phi}>-0.95$ for illustration.
It is then easy to see that $\widehat w_\phi(\widehat\Omega_\phi=0.7)>-0.95$
is a sufficient condition for ($w_{\phi}^{(0)}>-0.95$, $\Omega_{de}^{(0)}>0.7$).
Equivalently,
\be
\Gamma(\epsilon=3/28)<77/20
\label{e16}
\ee
Listed in Table I are the constraints on parameters of typical potentials by Eq.(\ref{e15}) and (\ref{e16}).

We note that for certain potentials $\widehat\Omega_\phi$
will tend to a maximum $\widehat\Omega_{max}$ smaller than 1 at last,
such as $U(\phi)=U_0e^{\phi^2/2}/\phi^n~(n>0,\phi>0)$ \cite{Brax:1999yv}.
These potentials always have a positive minimum $U_{min}$ at a finite $\phi$.
According to Eq.(\ref{e7}),
as the potential rolls to $U_{min}$,
$\eta=\epsilon\Gamma$ will tend to a nonzero minimum $\eta_{min}$ with
$\Gamma\rightarrow\infty$ and $\epsilon\rightarrow0$.
In this case, Eq.(\ref{e15}) and (\ref{e16}) are still valid though $\widehat\Omega_{\phi max}$
may be smaller than $0.7$.

\section{Discussions}
We have gotten stable fixed points $\widehat w_\phi$ and $\widehat\Omega_\phi$ from the full tracker equation, and
shown that they are good approximations for $w_\phi$ and $\Omega_\phi$ even in the $\Omega_\phi$ dominating period.
$\widehat w_\phi$ and $\widehat\Omega_\phi$ are analytic functions of $U(\phi)$.
The relation of $\widehat w_\phi\sim\widehat\Omega_\phi$ thus is gotten from the parametric functions
$\widehat w_\phi(\phi)$ and $\widehat\Omega_\phi(\phi)$,
which is also a good approximation to the relation of $w_\phi\sim\Omega_\phi$.

Formally, functions of $\widehat{w}_\phi$ and $\widehat{\Omega}_\phi$
with respect to expansion factor $a$ can also be obtained.
Substituting Eq.(\ref{e6}),(\ref{e7}) into the equation
\be
\frac{d\Omega_\phi}{d\ln a}=-3w_\phi\Omega_\phi(1-\Omega_\phi)
\ee
one gets the function of the field $\phi$ with respect to $a$ upon integration.
For example, for $U=U_0/\phi^2~~(\phi>0)$ one has:
\be
\phi(a)=\frac{\sqrt{14}}{3\sqrt{5}}(120a^3+49a^6)^{1/4}
\ee
where we have set present $\widehat \Omega_\phi^{(0)}=0.7$ and $a_0=1$.
Substituting $\phi(a)$ into Eq.(\ref{e6}) and Eq.(\ref{e7}) one gets:
\ba
\widehat\Omega_\phi&=&\frac{7}{60}(\sqrt{120a^3+49a^6}-7a^3)\nonumber\\
\widehat w_\phi&=&-\frac{1}{2}-\frac{7}{2\sqrt{120a^{-3}+49}}
\ea
For most potentials, it is not easy to get explicit functions of
$\phi(a)$, $\widehat w_\phi(a)$ and $\widehat\Omega_\phi(a)$.

The critical points $\widehat w_\phi$ and $\widehat\Omega_\phi$
can also be used to constrain the potential of quintessence directly
from observational conditions on ($w_{de},\Omega_{de}$).
We have adopted two conditions on present ($w_{de}^{(0)},\Omega_{de}^{(0)}$) for illustration.
Further astronomical observations will yield more properties of dark energy.
It may give conditions on ($w_{de},\Omega_{de}$) at other redshifts,
or even the exact shape of the $w_{de}\sim\Omega_{de}$ relation.
In that case, our method can be still usable to constrain the potential and
study the properties of the quintessence models that are fit with observations directly.

\begin{figure}[t]
\includegraphics[width=8cm,height=5cm]{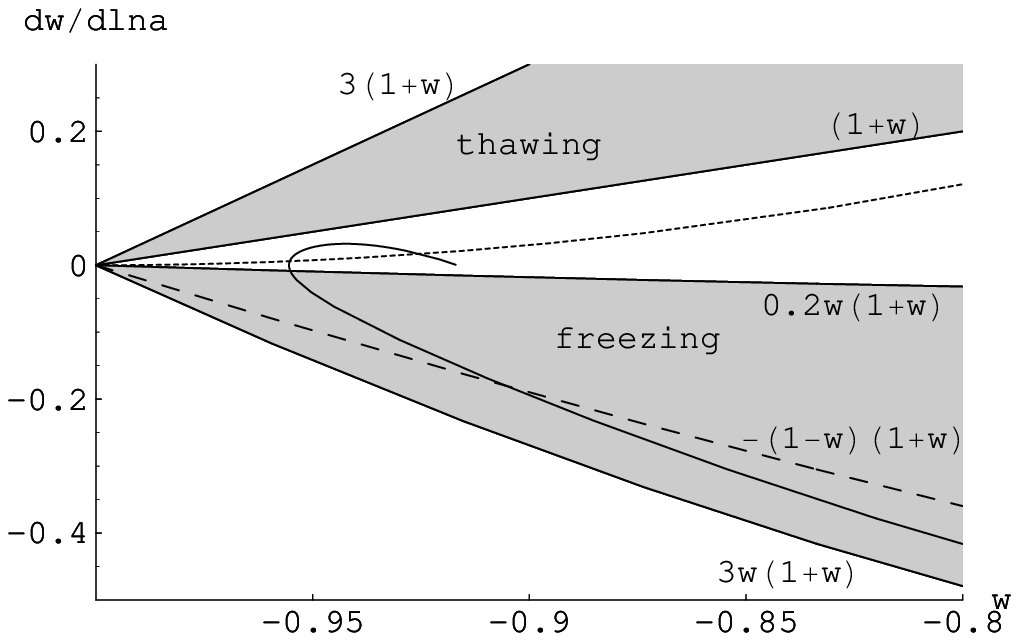}
\caption{The curve of the quintessence model with $U(\phi)=U_0(e^{-\phi/2}+e^{-20\phi})$ in the $w'-w$ phase space.
This curve crosses the the boundary for thawing and freezing fields \cite{Caldwell:2005tm} and the lower bound $w'=-(1-w)(1+w)$
in \cite{Scherrer:2005je,Chiba:2005tj}.}
\label{f2}
\end{figure}

In this paper, we have only studied the case that $w_\phi$~(and $\widehat w_\phi)$ keeps on decreasing monotonously,
from which the inequality (\ref{e12}) is obtained.
In fact, there are quintessence models in which $w_\phi$ is increasing at present.
One example is the case with $U(\phi)=U_0(e^{-\phi/2}+e^{-20\phi})$ \cite{Barreiro:1999zs}.
In this type of models,
$w_\phi$ will decrease to a minimum close to $-1$ and then begin to increase.
So the boundary for thawing and freezing fields in \cite{Caldwell:2005tm} will be crossed, as shown in Fig. \ref{f2}.
It can be shown that when
$$\frac{d\ln{(\Gamma-1)}}{d\ln{U}}>\frac{3}{2\epsilon}$$
$\widehat w_\phi$ will be increasing, so will be $w_\phi$.
It requires a rapid decrease of $\Gamma$.
As $\Gamma$ at early times must be close to 1 to get enough tracking,
usually there is a rapid increase of $\Gamma$ at recent times.
In this case the lower bound $w'>-(1-w)(1+w)$ for quintessence models \cite{Scherrer:2005je,Chiba:2005tj}
 may be crossed too.
It is because $w=(w_b-2\Gamma+2)/(2\Gamma-1)$ will no longer be larger than $w_\phi$
if the increase of $\Gamma$ is too fast.
It can be seen in Fig. \ref{f2} that the line of $w'\sim w$ with the double exponential potential
is very close to the strict lower bound $w'>3w(1+w)$ given in \cite{Caldwell:2005tm}.
For this type of potential, as $w_\phi$ and $\widehat w_\phi$ are increasing,
there is an inequality similar to (\ref{e12}):
\be
\widehat w_\phi>w_\phi,~~~~\widehat \Omega_\phi>\Omega_\phi
\ee
This inequality can be used to constrain $U(\phi)$ from conditions on ($w_{de},\Omega_{de}$).
The methods used in this paper can also be extended to Phantom and K-essence models.

\begin{acknowledgments}
This work is supported in part by the National Science Foundation of China (10425525).
\end{acknowledgments}

\end{document}